\documentclass[aps,twocolumn,nofootinbib,superscriptaddress,floatfix]{revtex4}

\usepackage{color,amsmath,amssymb,graphicx,latexsym,subfigure}

\usepackage{multirow}
\usepackage{ulem}

\usepackage{float}
\usepackage{booktabs}
\usepackage{threeparttable}
\usepackage[colorlinks,linkcolor=green, anchorcolor=green,citecolor=green]{hyperref}
\usepackage{ulem,soul}
\usepackage{xcolor}

\newcommand{\be}{\begin{equation}}
\newcommand{\ee}{\end{equation}}
\newcommand{\ba}{\begin{eqnarray}}
\newcommand{\ea}{\end{eqnarray}}

\newcommand{\gsim}{\mathrel{\hbox{\rlap{\lower.55ex \hbox {$\sim$}}
			\kern-.3em \raise.4ex \hbox{$>$}}}}
\newcommand{\lsim}{\mathrel{\hbox{\rlap{\lower.55ex \hbox {$\sim$}}
			\kern-.3em \raise.4ex \hbox{$<$}}}}

\providecommand{\kwd}[1]
{
  \small	
  \textbf{\textbf{Keywords:}} #1
}

\hypersetup{colorlinks=true,
	breaklinks=true,
	pdfstartview=Fit,
	linkcolor=blue,
	citecolor=blue,
	urlcolor=blue}

\begin{document}

\title{Quintom cosmology and modified gravity after DESI 2024}

\author{Yuhang Yang}
\affiliation{Department of Astronomy, School of Physical Sciences, University of Science and Technology of China, Hefei 230026, China}
\affiliation{CAS Key Laboratory for Researches in Galaxies and Cosmology, School of Astronomy and Space Science, University of Science and Technology of China, Hefei 230026, China}
\affiliation{Deep Space Exploration Laboratory, Hefei 230088, China}

\author{Xin Ren}
\affiliation{Department of Astronomy, School of Physical Sciences, University of Science and Technology of China, Hefei 230026, China}
\affiliation{CAS Key Laboratory for Researches in Galaxies and Cosmology, School of Astronomy and Space Science, University of Science and Technology of China, Hefei 230026, China}
\affiliation{Deep Space Exploration Laboratory, Hefei 230088, China}
\affiliation{Department of Physics, Tokyo Institute of Technology, Tokyo 152-8551, Japan}

\author{Qingqing Wang}
\affiliation{Department of Astronomy, School of Physical Sciences, University of Science and Technology of China, Hefei 230026, China}
\affiliation{CAS Key Laboratory for Researches in Galaxies and Cosmology, School of Astronomy and Space Science, University of Science and Technology of China, Hefei 230026, China}
\affiliation{Deep Space Exploration Laboratory, Hefei 230088, China}

\author{Zhiyu Lu}
\affiliation{Department of Astronomy, School of Physical Sciences, University of Science and Technology of China, Hefei 230026, China}
\affiliation{CAS Key Laboratory for Researches in Galaxies and Cosmology, School of Astronomy and Space Science, University of Science and Technology of China, Hefei 230026, China}
\affiliation{Deep Space Exploration Laboratory, Hefei 230088, China}

\author{Dongdong Zhang}
\affiliation{Department of Astronomy, School of Physical Sciences, University of Science and Technology of China, Hefei 230026, China}
\affiliation{CAS Key Laboratory for Researches in Galaxies and Cosmology, School of Astronomy and Space Science, University of Science and Technology of China, Hefei 230026, China}
\affiliation{Deep Space Exploration Laboratory, Hefei 230088, China}
\affiliation{Kavli IPMU (WPI), UTIAS, The University of Tokyo, Kashiwa, Chiba 277-8583, Japan}

\author{\\ Yi-Fu Cai} \email{yifucai@ustc.edu.cn}
\affiliation{Department of Astronomy, School of Physical Sciences, University of Science and Technology of China, Hefei 230026, China}
\affiliation{CAS Key Laboratory for Researches in Galaxies and Cosmology, School of Astronomy and Space Science, University of Science and Technology of China, Hefei 230026, China}
\affiliation{Deep Space Exploration Laboratory, Hefei 230088, China}

\author{Emmanuel N. Saridakis} \email{msaridak@noa.gr}
\affiliation{National Observatory of Athens, Lofos Nymfon 11852, Greece}
\affiliation{CAS Key Laboratory for Researches in Galaxies and Cosmology, School of Astronomy and Space Science, University of Science and Technology of China, Hefei 230026, China}
\affiliation{Departamento de Matem\'{a}ticas, Universidad Cat\'{o}lica del Norte, Avda. Angamos 0610, Casilla 1280, Antofagasta, Chile}

%\author{Xinmin Zhang} \email{xmzhang@ihep.ac.cn}
%\affiliation{Theoretical Physics Division, Institute of High Energy Physics, Chinese Academy of Sciences, 19B Yuquan Road, Shijingshan District, Beijing 100049, China}
%\affiliation{School of Physics, University of Chinese Academy of Sciences, Beijing 100049, China}

\begin{abstract}
We reconstruct the cosmological background evolution under the scenario of dynamical dark energy through the Gaussian process approach, using the latest Dark Energy Spectroscopic Instrument (DESI) baryon acoustic oscillations (BAO) combined with other observations.
Our results reveal that the reconstructed dark-energy equation-of-state (EoS) parameter $w(z)$ exhibits the so-called quintom-B behavior, crossing $-1$ from phantom to quintessence regime as the universe expands. We investigate under what situation this type of evolution could be achieved from the perspectives of field theories and modified gravity. In particular, we reconstruct the corresponding actions for $f(R)$, $f(T)$, and $f(Q)$ gravity, respectively. We explicitly show that, certain modified gravity can exhibit the quintom dynamics and fit the recent DESI data efficiently, and for all cases the quadratic deviation from the $\Lambda$CDM scenario is mildly favored.
\\

\par\kwd{DESI, dark energy, quintom cosmology, modified gravity}
\end{abstract}

\date{\today}

\maketitle
%\section{Introduction} \label{sec:intro}
%Basic cosmological background and introduction of DESI

\section{Introduction}
With the era of precision cosmology (such as the latest data release of Dark Energy Spectroscopic Instrument (DESI) baryon acoustic oscillations (BAO) \cite{DESI:2024mwx}), our understanding on the evolution of the universe has greatly advanced. Astonishingly, the High Redshift Supernova Team \cite{SupernovaSearchTeam:1998fmf} and the Supernova Cosmology Project \cite{SupernovaCosmologyProject:1998vns} discovered that distant Type Ia supernovae (SN Ia) were accelerating away at an increasing pace, following which further evidence from the Cosmic Microwave Background (CMB) \cite{Planck:2018vyg}, BAO \cite{BOSS:2016wmc,eBOSS:2020yzd,Mehta:2012hh}, and large-scale structure survey \cite{DAmico:2019fhj,Ivanov:2019pdj,Chuang:2013hya} confirmed the accelerating expansion as well. This led to the concept of dark energy, responsible for such a phenomenon, but the underlying nature remains mysterious.
Facing to the aforementioned phenomenon, there are growing interests in various cosmological models. Despite of the simplest version of the cosmological constant $\Lambda$, there are many other candidate scenarios, namely dynamical dark energy models \cite{Copeland:2006wr, Gubitosi:2012hu, Creminelli:2017sry,Teng:2021cvy}.
Some implementations of dynamical dark energy are known as scalar-field models, including quintessence \cite{Ratra:1987rm, Wetterich:1987fm}, phantom \cite{Caldwell:1999ew}, quintom \cite{Feng:2004ad}, K-essence \cite{Armendariz-Picon:2000ulo, Malquarti:2003nn} and so on.
The feature shared by all these models is a time-evolving equation-of-state (EoS) $w$. For quintessence, the value of $w$ is always larger than $-1$, while for phantom $w<-1$. Meanwhile, $w$ can cross $-1$, thereby enabling the description of a broader range of cosmological evolution in quintom cosmology \cite{Xia:2004rw, Xia:2005ge, Zhao:2005vj, Guo:2006pc}. To be specific, in the  quintom-A scenario $w$ is arranged to evolve from above $-1$ at early times to below $-1$ at late times; while, in quintom-B $w$ changes from the phantom phase to the quintessence phase as the universe expands. Note that, in general the realization of quintom-B is challenging when compared to quintom-A \cite{Cai:2006dm, Cai:2009zp}.

It is worth mentioning that, some observations put hints on an existence of the negative-valued effective energy density of dark energy at high redshifts \cite{Wang:2018fng, Dutta:2018vmq, Visinelli:2019qqu, Vagnozzi:2019ezj, Abdalla:2022yfr, Adil:2023ara, Menci:2024rbq, Malekjani:2023ple}, which poses a challenge to the scalar field theory of dark energy, as it violates the null energy condition \cite{Buniy:2005vh, Qiu:2007fd, Cai:2009zp}. Theoretically, modified gravity \cite{CANTATA:2021ktz} can be a framework to provide an alternative explanation for the above issue. Particularly, in modified gravity the additional terms relative to general relativity can behave as a component with the dynamical EoS, and thus can serve as an effective form of dynamical dark energy.
One can develop curvature-based extended gravitational theories, such as $f(R)$ gravity \cite{Starobinsky:1980te, Capozziello:2002rd, DeFelice:2010aj,Nojiri:2003ft,Nojiri:2010wj}. Modified gravity theories can also be constructed based on other geometric gravity equivalent to general relativity. Starting from the torsion-based Teleparallel Equivalent of General Relativity, one can extend it to $f(T)$ gravity \cite{Cai:2015emx, Krssak:2015oua, Krssak:2018ywd, Bahamonde:2021gfp}. The extensions of Symmetric Teleparallel Equivalent of General Relativity based on non-metricity leads to $f(Q)$ gravity \cite{BeltranJimenez:2017tkd, Heisenberg:2023lru}. These theories have been widely studied in cosmological frameworks \cite{BeltranJimenez:2019tme, Cai:2011tc, Clifton:2011jh, Nojiri:2017ncd}.

% DESI
Confronted with the landscape of theoretical upsurge such as the physical meaning of dark energy, and the gravitational descriptions underpinning the geometry of the universe, there is an urgent need for observational guidance to steer the course of theoretical development. BAO data act as a powerful tool for probing cosmic distances, and play a pivotal role in the study of dark energy properties. Previous works had found implications of dynamical dark energy: $3.5\sigma$ evidence by using Bayesian Method with the data from SDSS DR7, BOSS and WiggleZ \cite{Zhao:2012aw, Zhao:2017cud, Colgain:2021pmf, Pogosian:2021mcs}. Recently, the release of DESI provided measurements of the transverse comoving distance and Hubble rate, showing a possible tension with respect to the $\Lambda$CDM scenario at the level of 3.9$\sigma$ \cite{DESI:2024mwx}. Combining the DESI data with CMB and Supernova, provides indications of a deviation from a cosmological constant in favor of dynamical dark energy in Ref. \cite{Cortes:2024lgw}.
Thus, confrontation with DESI data has attracted the interest of the community, suggesting interacting dark energy \cite{Giare:2024smz}, quintessence scalar fields \cite{Berghaus:2024kra,Tada:2024znt}, dark radiation \cite{Allali:2024cji}, and other scenarios beyond $\Lambda$CDM paradigm \cite{Gomez-Valent:2024tdb,Wang:2024hks, Colgain:2024xqj, Carloni:2024zpl, Wang:2024rjd, Yin:2024hba, Luongo:2024fww}.

In this work, we take full advantage of the most recent DESI data to reconstruct the dynamic evolution of our universe via the model-independent Gaussian process. We explain the quintom behavior of $w(z)$ within the framework of modifications of gravity, including $f(R)$, $f(T)$, and $f(Q)$ theories,  then reconstruct the corresponding involved unknown function.

%DATA COMPONENT
\section{Dynamical evolution and quintom cosmology}
BAO measurements are conducted across various redshift intervals, thereby enabling the imposition of constraints upon the cosmological parameters that regulate the distance-redshift relationship.
The DESI BAO data includes tracers luminous red galaxy (LRG), emission line galaxies (ELG) and the Lyman-$\alpha$ forest (Ly$\alpha$ QSO) in a redshift range $0.1\leq z \leq 4.2$ \cite{DESI:2024uvr,DESI:2024lzq}.
The preliminary data includes quantities of $D_{\rm {M}}(z)/r_{\rm {d}}$, $ D_{\rm {H}}(z)/r_{\rm {d}}$ and $D_{\rm {V}}(z)/r_{\rm {d}}$ within 7 distinct redshift bins. Here $r_{d}$ is the drag-epoch sound horizon and the transverse comoving distance $D_{\rm{M}}(z) = r_{\rm {d}}/\Delta\theta $, equivalent distance $D_{\rm{H}}(z) = c/H(z)$ and angle-average distance $D_{\rm{V}}(z) = (zD_{\rm{M}}^2(z)D_{\rm{H}}(z))^{1/3}$.
For later reconstruction we use the $5$ $ D_{\rm{H}}(z)/r_{\rm {d}}$ data and assume no derivation from $\Lambda$CDM at high redshift, thus imposing
$r_{\rm {d}}=147.09 \pm 0.26$ Mpc \cite{Planck:2018vyg} obtained from CMB to directly calibrating the BAO standard ruler.

To investigate the impact of the DESI data on the dark-energy EoS parameter, we consider three scenarios:
in the first case, we exclusively utilize the
%comoving
distance data from DESI to reconstruct the evolution of the Hubble parameter with redshift.
As a control sample, the second group consists solely of data from SDSS and WiggleZ, which serves to verify whether the results from DESI indeed provide stronger evidence for models featuring dynamical dark energy. For the third scenario, we combine the DESI data with complementary datasets \cite{Mukherjee:2021ggf, Wu:2022fmr, eBOSS:2020yzd, Wang:2024qan} from SDSS and WiggleZ. All the samples we used including five DESI data and previous BAO (P-BAO) data are listed in the Supplementary materials Section~A. The covariance matrix of all the data points are assumed to be diagonal. To validate this assumption, we combine independent datasets: WiggleZ \cite{Blake:2012pj}, BOSS DR12 \cite{BOSS:2016wmc}, and eBOSS DR16 \cite{eBOSS:2020hur,eBOSS:2020lta,eBOSS:2020yql,eBOSS:2020gbb,eBOSS:2020uxp,eBOSS:2020tmo}. The reconstruction result of $w$ exhibit comparable behavior, differing by approximately 15\% from subsequent results, which indicates that this assumption is sufficiently robust.

In order to reconstruct the history of cosmic dynamics evolution from the BAO data, we perform a model-independent reconstruction of the Hubble parameter by using the Gaussian process. The Gaussian process is a stochastic procedure to acquire a Gaussian distribution over functions from observational data \cite{Shafieloo:2012ht}, which has been widely used for the function reconstruction in cosmology \cite{Cai:2019bdh, Ren:2021tfi, Aljaf:2020eqh,  LeviSaid:2021yat, Bonilla:2021dql, Bernardo:2021qhu, Ren:2022aeo,
Elizalde:2022rss, Liu:2023agr, Fortunato:2023ypc, Yang:2024tkw}. The distribution of the function at different redshifts is related by the covariance function with hyperparameters. We reconstruct the evolution function of $H(z)$ and its derivative through Gaussian Process in Python (GAPP) based on the exponential covariance function $k\left(x, x^{\prime}\right)=\sigma_{\rm{f}}^{2}
e^{-\left(x-x^{\prime}\right)^{2}/(2 l^{2})}$, where the $\sigma_{\rm{f}}$ and $l$ are the hyperparameters \cite{Seikel:2012uu}.

\begin{figure}[htbp]
    \centering
    \includegraphics[width=\columnwidth]{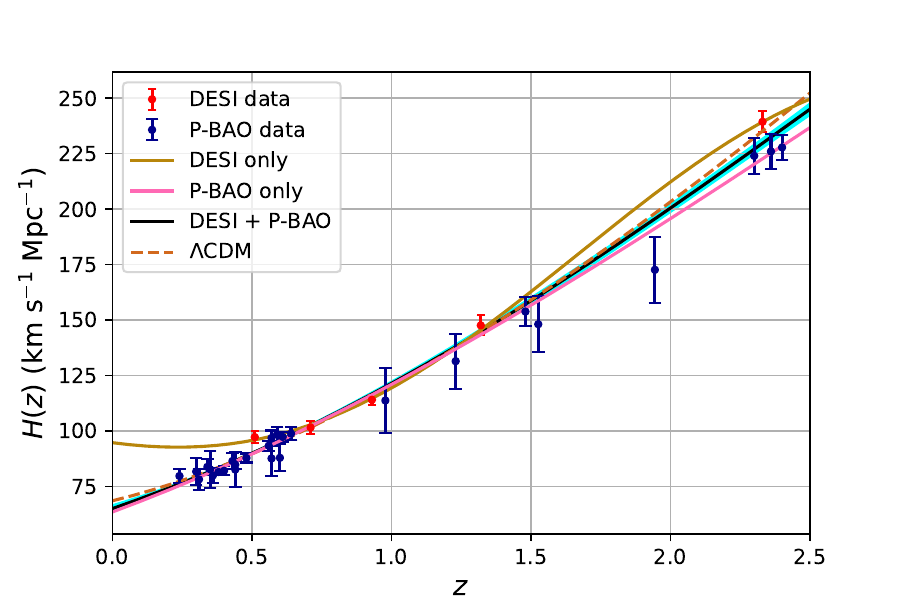}
    \caption{{\it{The reconstructed $H(z)$   arising from DESI and P-BAO data through Gaussian process, without imposing the value of $H_0$. The black curve denotes the mean value, while the light blue shaded zones indicate the allowed regions at $1\sigma$ confidence level for DESI + P-BAO. The dashed line corresponds to
      the $\Lambda$CDM scenario with the best fit value $H_0=68.52\pm 0.62$ $\rm km$ $\rm s^{-1}Mpc^{-1}$ of Ref. \cite{DESI:2024mwx}, while the mean value of the reconstructed $H(z)$ from DESI only or P-BAO only are additionally presented with the brown   and pink curves respectively.}}}
    \label{fig:H(z) data}
\end{figure}

By applying the GAPP steps, we obtain the reconstructed $H(z)$ function which is depicted in Fig.~\ref{fig:H(z) data}. The black curve denotes the mean value of the reconstruction by using DESI and P-BAO data, while the light blue shaded zones indicate the allowed regions at $1\sigma$ confidence level. Furthermore, the $\Lambda$CDM model has been depicted with the dash line, imposing the best fit $H_0=68.52 \pm 0.62$ $\rm km$ $\rm s^{-1}Mpc^{-1}$ in \cite{DESI:2024mwx}. One can read that, at low redshift it can fit the reconstruction result by using DESI and P-BAO data well, but at high redshift the differences are statistically significant, as the $\Lambda$CDM results are higher than those derived from reconstruction. Meanwhile, the mean values of the reconstructed $H(z)$ by using DESI or P-BAO only is also shown in the figure. We find that the value $H_0^{\mathrm{DESI}}=94.22 \pm 13.81$ $\rm km$ $\rm s^{-1}Mpc^{-1}$ which results from only DESI is too high to fit CMB or SNIa observations. This suggests that due to the limited number of BAO data points from DESI, there is an absence of information for low redshift bins.
%Relying solely on DESI is insufficient to achieve a complete reconstruction result by GAPP, hence in the later gravitational reconstruction we will not consider the DESI only situation.
Moreover, we also acquire the $H_0$ values from the reconstruction processes for other two cases, which are $H_0^{\mathrm{P-BAO}}=63.08 \pm 2.94$  $\rm km$ $\rm s^{-1}Mpc^{-1}$, $H_0^{\mathrm{DESI+P-BAO}}=64.64 \pm 2.52$ $\rm km$ $\rm s^{-1}Mpc^{-1}$, respectively.

Surprisingly, we notice that the DESI data point around $z=0.51$ is much higher than the range of $1\sigma$ allowed by the reconstruction result. Actually, the DESI data near $z=0.51$ is $2.44\sigma$ away from the P-BAO only result and $2.42\sigma$ away from DESI + P-BAO.
This unexpected phenomenon is also mentioned in Refs. \cite{Colgain:2024xqj, Giare:2024smz}, and if it indeed arises from systematics, a possible explanation would be statistical fluctuations. Thus, in the future we may need more observational data at $z=0.51$ to extract  more precise results.

We then use the $H(z)$ function presented above to reconstruct the dark-energy EoS. Following the Friedmann equations, one can easily define the dark-energy EoS as
\begin{equation}
    w=\frac{-2\dot{H}-3H^2-p_{\rm{m}}}{3H^2-\rho_{\rm{m}}},
\end{equation}
where $c\equiv 8\pi G \equiv 1$ is adopted. $\rho_{\rm{m}}$ and $p_{\rm{m}}$ are the energy density and pressure of the matter sector (baryonic plus cold dark matter), assuming it to be a perfect fluid. One can easily find  $\rho_{\rm{m}}=3H_0^3\Omega_{\rm{m}0}(1+z)^3$ by using the continuity equation of matter $\dot{\rho}_{\rm{m}}+3H(\rho_{\rm{m}}+p_{\rm{m}})=0$, where $\Omega_{\rm{m}0}=0.3153 \pm 0.0073$ is the present value of the matter density parameter measured by Planck \cite{Planck:2018vyg}.

\begin{figure*}[htbp]
	\centering
	\subfigure{
		\begin{minipage}[t]{0.325\linewidth}
			\centering
			\includegraphics[width=\textwidth]{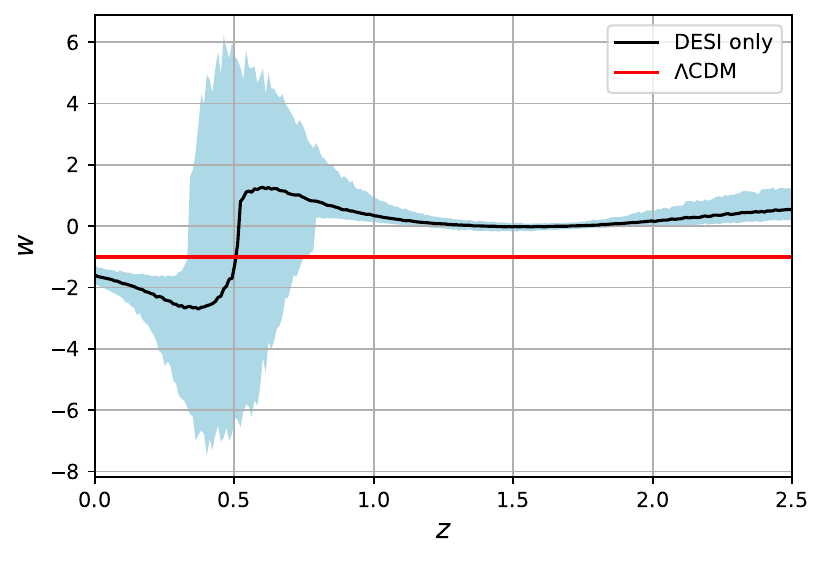}
            \label{fig:wde_z_desi}
		\end{minipage}
	}%
	\subfigure{
		\begin{minipage}[t]{0.325\linewidth}
			\centering
			\includegraphics[width=\textwidth]{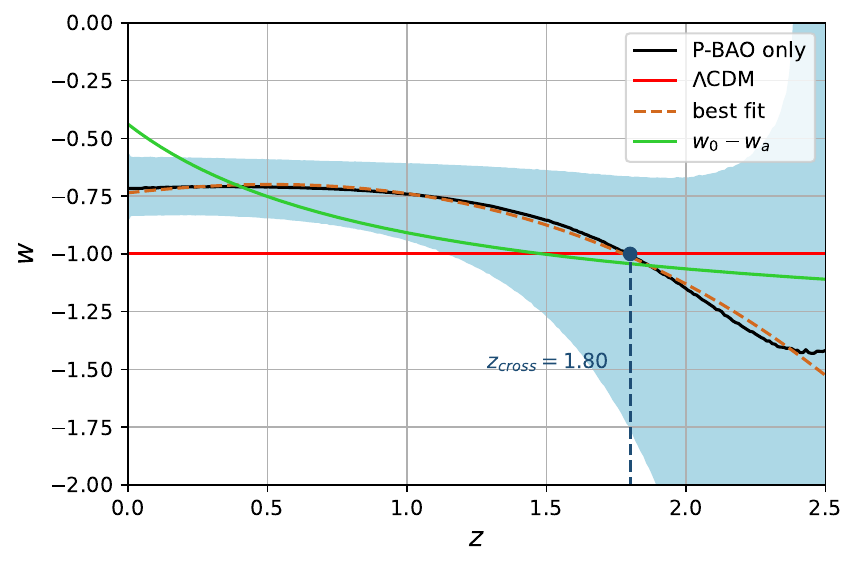}
            \label{fig:wde_z_bao}
		\end{minipage}
	}%
	\subfigure{
		\begin{minipage}[t]{0.325\linewidth}
			\centering
			\includegraphics[width=\textwidth]{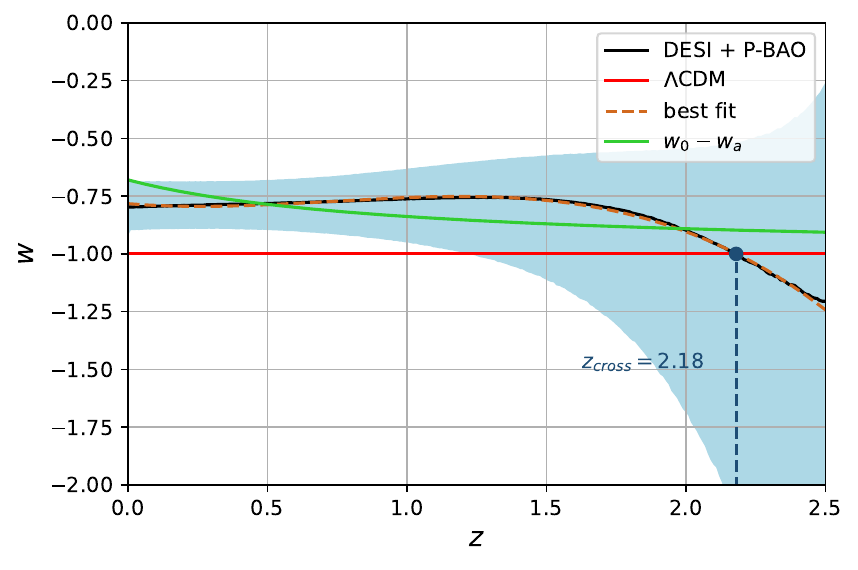}
            \label{fig:wde_z_desi_bao}
		\end{minipage}
	}%
	\centering
	\caption{{\it{The reconstructed dark-energy EoS parameter $w(z)$ for different datasets. The black curve denotes the mean value, while the light blue shaded zones indicate the allowed regions at $1\sigma$ confidence level.
  Additionally, we depict the best fit function $w(z)=a+bz+cz^2+dz^3$, for P-BAO only data  (where $a=-0.73,b=0.13,c=-0.10,d=-0.03$) and for DESI + P-BAO data (where $a=-0.78,b=-0.10,c=0.23,d=-0.11$). Furthermore, we depict the best-fit curves corresponding to $w_0$-$w_a$ parametrization by green curves.} Finally, the red curve corresponds to the
 $\Lambda$CDM scenario in which $w=-1$. $z_{\rm{cross}}$ marks the point in which the phantom divide is crossed. }}
	\label{fig:wde_z}
\end{figure*}

The reconstructed $w(z)$ for different data set is shown in Fig.~\ref{fig:wde_z}. It is worth emphasizing that our reconstruction method, namely the Gaussian process, is model-independent, which implies that we do not need to parameterize the evolution of $w$ as priors. Hence, we can obtain the evolution characteristics and behavior of $w$ in a model-independent way. The mean values of $w(z)$ given by the three sets of data, all tend towards dynamical evolution. The results show that $w$ has a tendency to cross zero for DESI data only, resulting from the divergence when the effective dark energy density crosses zero. For the result from P-BAO or the combined data, $w$ exhibits a quintom-B behavior, which implies that it can cross $-1$ from the phantom phase to the quintessence phase. Further, we calculate the confidence of the quintom-B dynamics using the Monte Carlo simulation and obtain results of $0.93 \sigma$ and $0.78\sigma$ for P-BAO only and DESI + P-BAO, which shall be better constrained by combining CMB and SN Ia data.
The crossing redshift, in which  $w$  crosses $-1$, is found to be $1.80$, $2.18$ for P-BAO only and DESI + P-BAO respectively, which indicate that the presence of DESI data can increase the value of $w$ at high redshifts since the value of $H$ at $z=2.33$ from DESI is also larger than other data at the same redshift. It is worth noting that a similar quintom-B behavior of dark energy has also been found in previous articles \cite{Cortes:2024lgw}, however the difference is that here we use BAO data to reconstruct $w$ in a model-independent way, while in that work they used SN Ia data to perform the Monte Carlo Markov Chain method by assuming the evolution of $w$. Additionally, we find a different value for the crossing $z$. Meanwhile the results also show that $\Lambda$CDM scenario is beyond the $1\sigma$ allowed regions at low redshifts for both P-BAO only and DESI + P-BAO.

Additionally, with  the green curve  in Fig.~\ref{fig:wde_z}  we depict the best-fit result of $w_0$-$w_a$ parametrization, namely $w=w_0+w_a(1-a)$ where $w_0, w_a$ are free parameters. It is evident that while the best fit of $w_0$-$w_a$ parametrization still falls within the reconstructed 1 $\sigma$ region, it deviates from the mean value, indicating that a simple parametrization of dark energy evolution using traditional $w_0$-$w_a$ may not be sufficient. Therefore, higher order terms beyond linear order need to be introduced. To fit the model-independent reconstruction result of $w$, we use the parametrization, namely
\begin{equation}
        w(z)=a+bz+cz^2+dz^3,
        \label{wparametrization}
\end{equation}
where $a,b,c,d$ are dimensionless parameters. The parameter values are presented in Table~\ref{table:best fit value wde}, while the best fit curves are also shown in Fig.~\ref{fig:wde_z}.

\begin{table}[htbp]
\centering

\caption{The $a,b,c$ and $d$   dimensionless parameter values for
best fitting, according to parametrization  (\ref{wparametrization}).}
\begin{tabular}{ccc}
\hline

Data&P-BAO& DESI + P-BAO \\
\hline
$a$ &$-0.73$&$-0.78$  \\
$b$ &0.13 &0.10  \\
$c$ &0.10&0.23 \\
$d$ &$-0.03$&$-0.11$ \\
\hline
\end{tabular}
\label{table:best fit value wde}
\end{table}

%data set: DESI, DESI + BAO, DESI + BAO + OHD?

%put H(z) figure and wde figure for different data set
It is worth emphasizing that according to the ``No-Go'' theorem, the EoS parameters of a single scalar field is forbidden to cross $-1$ \cite{Cai:2009zp, Hu:2004kh, Kunz:2006wc}. Therefore, this reconstruction results pose a significant challenge to the single scalar field dark energy model. The quintom model can be realized through various theories such as two scalar fields \cite{Guo:2004fq, Zhang:2005eg}, spinor fields \cite{Alimohammadi:2008mh}, string theory \cite{Cai:2007gs}, DHOST \cite{Langlois:2017mxy, Langlois:2018jdg} and Horndeski \cite{Horndeski:1974wa}, more details are available in \cite{Cai:2009zp}. Due to the ``No-Go'' theorem, the explicit construction of the quintom scenario is more complex than that of other dynamical dark energy models. The realization of the quintom scenario requires a non-zero derivative of $w$ near the crossing point. Also both the background and perturbations of scalar field must be stable and cross the boundary smoothly.

Meanwhile, the quintom model is widely used in the early universe. In a bouncing universe scenario, the universe initially contracts to a non-vanishing minimal radius before entering a subsequent phase of expansion. Following the bounce, as the universe transits into the hot Big Bang era, the EoS must shift from $w < -1$ to $w > -1$. This transition is characteristic of a quintom scenario \cite{Cai:2007qw, Cai:2007zv}. The quintom dynamics can also be utilized to realize cyclic cosmology \cite{Xiong:2008ic} and emergent universe \cite{Cai:2012yf, Cai:2013rna, Ilyas:2020zcb}, potentially providing a solution to the singularity problem in the Big Bang cosmology.

One typical way to obtain a realization of the quintom-like phenomenon is within two scalar fields theory, if we combine one quintessence scalar field $\phi$ and one phantom scalar field $\sigma$. In such a case, the EoS parameter of quintom dark energy $w_{\rm {q}}$ can be written as
\begin{align}
    w_{\rm {q}}=\frac{p_\phi+p_\sigma}{\rho_\phi+\rho_\sigma}=\frac{\frac{1}{2}\dot{\phi}^{2} - V_\phi(\phi)-\frac{1}{2}\dot{\sigma}^{2} - V_\sigma(\sigma)}{\frac{1}{2}\dot{\phi}^{2} + V_\phi(\phi)-\frac{1}{2}\dot{\sigma}^{2} + V_\sigma(\sigma) }~,
\end{align}
where $V_\phi(\phi)$, $V_\sigma(\sigma)$ are the potentials for each scalar field respectively. However, the appropriate potentials and initial conditions to realize the quintom behavior is quiet difficult to be chosen. Nevertheless, since phantom scalar fields may exhibit problems at the quantum level \cite{Vikman:2004dc,Cline:2003gs}, it would be more natural and simpler to explain the quintom behavior within modified gravity framework.

%\section{Gravitational reconstruction}
%\label{sec:reconstruction}

\section{Gravitational reconstruction}
For the gravitational reconstruction, we consider metric-affine gravity \cite{Hehl:1994ue}, describing gravity with a metric and a general affine connection. Such a general formulation can reduce to $f(R)$, $f(T)$, and $f(Q)$ gravity under certain conditions, based only on curvature, torsion or non-metricity respectively. These three metric-affine modified gravity theories constitute the geometric trinity of gravity.
The action for curvature $f(R)$ gravity, torsional $f(T)$ gravity and non-metric $f(Q)$ gravity can be uniformly expressed as \cite{DeFelice:2010aj,Cai:2015emx,BeltranJimenez:2017tkd}
\begin{align}
    S = \int d^4 x \sqrt{-g} \left[ \frac{1}{2} f(X) + \mathcal{L}_{\rm{m}} \right] ~,
\end{align}
where $X$ represents $R,T$ or $Q$, with $R,T,Q$ the Ricci scalar, torsion scalar and non-metricity scalar, $\mathcal{L}_{\rm{m}}$ represents the matter Lagrangian density respectively.
To apply these modified gravity theories in a cosmological framework, we consider the isotropic and homogeneous flat  Friedmann-Robertson-Walker (FRW) metric ${\rm d}s^2={\rm d}t^2-a(t)^2({{\rm d}r^2}+r^2{\rm d}\theta^2+r^2\sin^2\theta {\rm d}\phi^2)$,
with $a(t)$ the scale factor. The modified Friedmann equations can be expressed effectively as
\begin{align}
\label{eq:frteq}
\begin{aligned}
    3H^2&=\rho_{\rm{m}}+\rho_{\rm{de}} ~, \\
    -2\dot{H}-3H^2&=p_{\rm{m}}+p_{\rm{de}} ~,
\end{aligned}
\end{align}
where $\rho_{\rm{m}}$ and $p_{\rm{m}}$ denote the energy density and pressure of matter, and the effective energy density $\rho_{\rm{de}}$ and pressure $p_{\rm{de}}$ are in terms of the gravitational modifications.

\begin{figure*}[htbp]
\centering
    \subfigure{
    \begin{minipage}[t]{0.45\linewidth}
    \centering
    \includegraphics[width=0.8\textwidth,height=4cm]{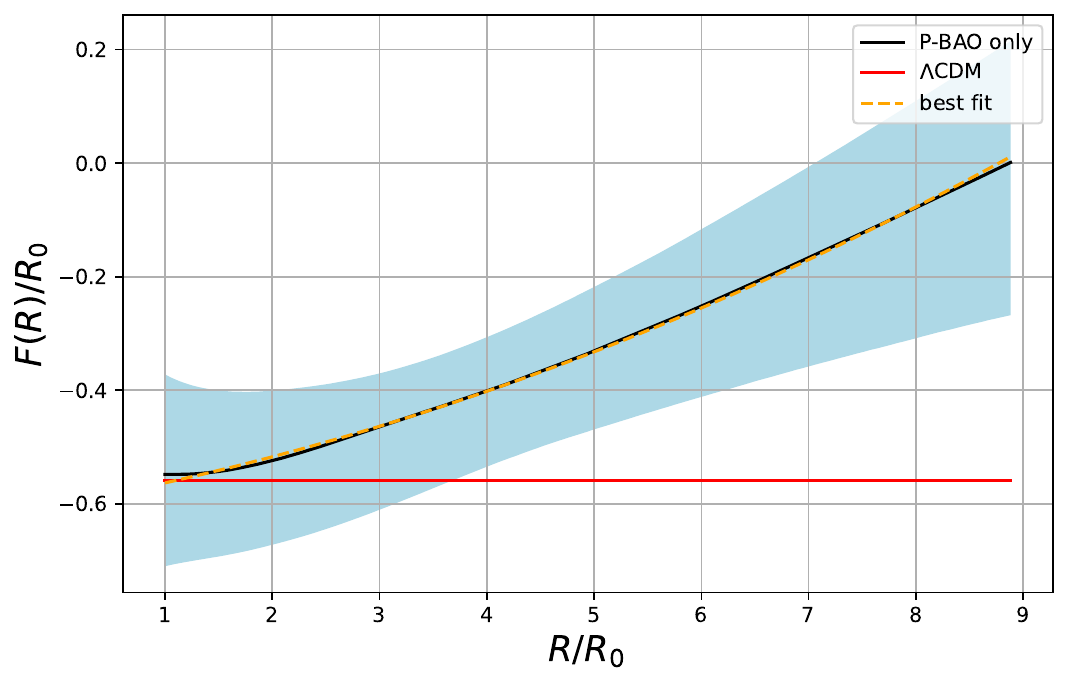}
    \label{fig:fR_R_bao}
    \end{minipage}
    }
    \subfigure{
    \begin{minipage}[t]{0.45\linewidth}
    \centering
    \includegraphics[width=0.8\textwidth,height=4cm]{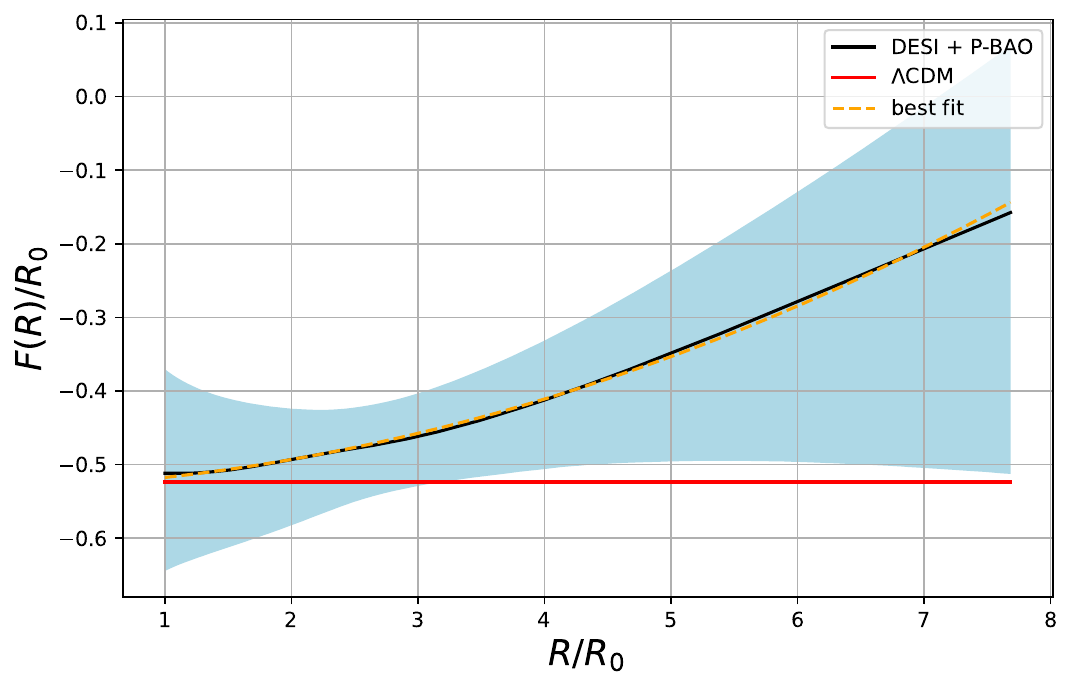}
    \label{fig:fR_R_desi_bao}
    \end{minipage}
    }
    \subfigure{
    \begin{minipage}[t]{0.45\linewidth}
    \centering
    \includegraphics[width=0.8\textwidth,height=4cm]{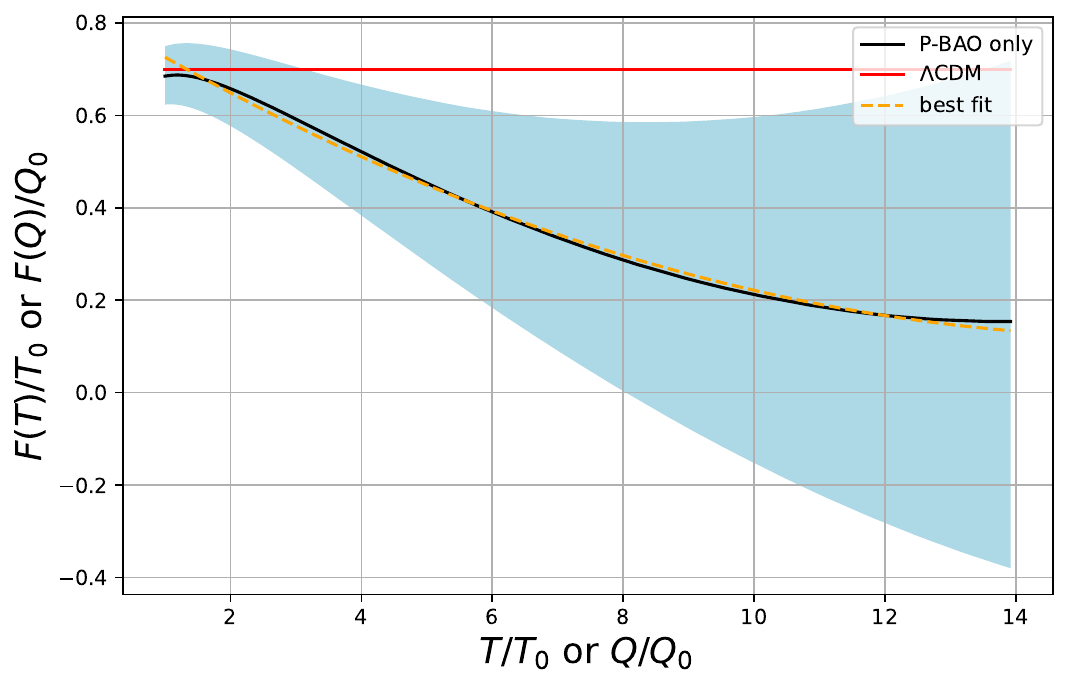}
    \label{fig:fQ_Q_bao}
    \end{minipage}
    }
    \subfigure{
    \begin{minipage}[t]{0.45\linewidth}
    \centering
    \includegraphics[width=0.8\textwidth,height=4cm]{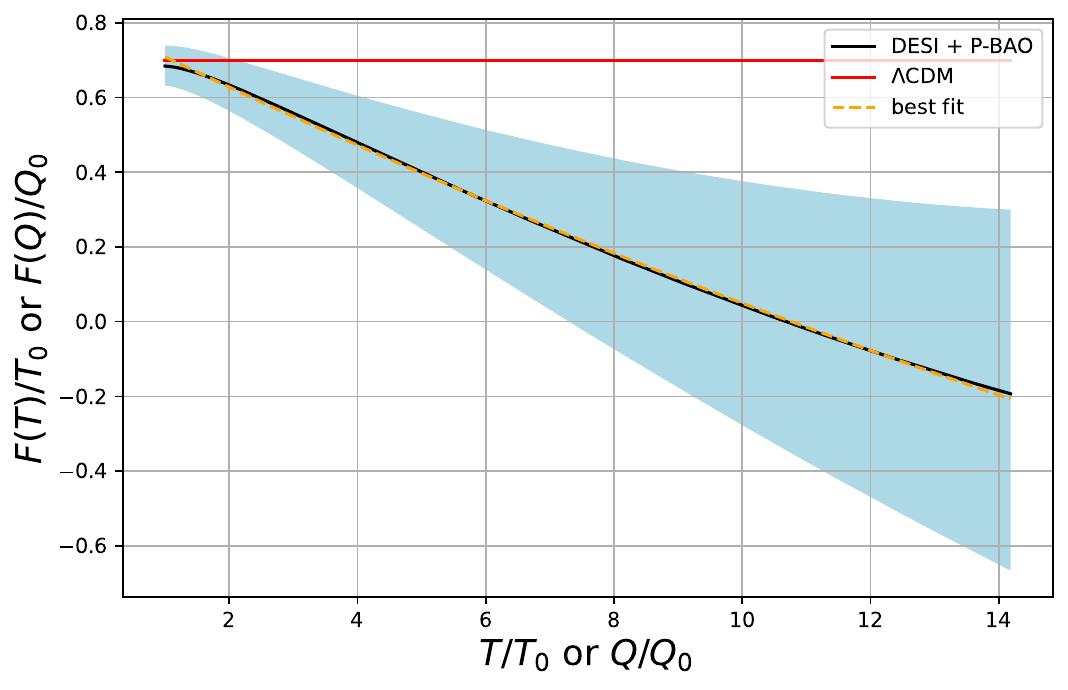}
    \label{fig:fQ_Q_desi_bao}
    \end{minipage}
    }
    \centering
    \caption{{\it{The reconstructed $F(X)$ for different datasets, with $F(X)=f(X)-X$, where $X$ represents $R,T$ or $Q$, and with $X_0$ the current value. The black curve denotes the mean value, while the light blue shaded zones indicate the allowed regions at $1\sigma$ confidence level. The upper panels show the result of $F(R)$ gravity, while the lower panels show the result of $F(T)$ or $F(Q)$ gravity (since they coincide at the background level for FRW geometry within the coincident gauge). We use the parametrization \eqref{FXpaeametrization}, i.e., $F(X)/X_0=A+BX/X_0+CX^2/X_0^2$, to fit the reconstruction result, with the parameter values shown in Table \ref{table:best fit value f} for P-BAO only and DESI + BAO data set, respectively. Additionally, the red line depicts the $\Lambda$CDM scenario, with $\Lambda^{\rm{P-BAO}}=0.7\times 3H_0^{2 \ \rm{P-BAO}}$ and $\Lambda^{\rm{DESI+P-BAO}}=0.7\times 3H_0^{2 \ \rm{DESI+P-BAO}}$.}}}
\label{fig:fX_X}
\end{figure*}

In $f(R)$ gravity, we have
\begin{equation}
\begin{aligned}
 \rho_{{\rm de},R}=&\frac{1}{f_R}\left[\frac{1}{2}(f-Rf_R)-3H\dot{R}f_{RR} \right] ~, \\
 p_{{\rm de},R}=&\frac{1}{f_R}(2H\dot{R}f_{RR}+\Ddot{R}f_{RR}) \\
 & +\frac{1}{f_R}\left[\dot{R}^2f_{RRR} -\frac{1}{2}(f-R f_R)\right] ~,
\end{aligned}
\end{equation}
where $R=-12H^2-6\dot{H}$ and $f_R={\rm d}f/{\rm d}R$, $f_{RR}={\rm d}^2f/{\rm d}R^2$, and accordingly the effective dark-energy EoS is $w\equiv p_{\rm{de},R}/\rho_{\rm{de},R}$.

Similarly, in $f(T)$ gravity, we have the torsional energy density and pressure as
\begin{equation}
\begin{aligned}
 \rho_{\rm{de},T} &= -\frac{1}{2}F +TF_T ~,  \\
 p_{\rm{de},T} &= \frac{F-TF_T+2T^2F_{TT}}{2+2F_T+4TF_{TT}} ~,
\end{aligned}
\end{equation}
where we have introduced $f(T)=T+F(T)$ for convenience, and with $T=-6H^2$, and thus the effective dark-energy EoS parameter is $w\equiv p_{\rm{de},T}/\rho_{\rm{de},T}$. For $f(Q)$ gravity within coincident gauge, in the FRW metric at the background level, where $Q=-6H^2$, the corresponding expressions can be obtained from the one of $f(T)$ gravity, with the replacement $T\rightarrow{Q}$.

Since we have reconstructed the evolution of the dark-energy EoS parameter from the data, and we have expressed it in terms of the modified gravity involved function, based on $H(z)$ and its derivative we can straightforwardly obtain the reconstruction of these functions too in a nearly model-independent way. The details are provided in the Supplementary materials Section~B.
Then, from the Supplementary materials Section~B we can reconstruct the evolution of $f(z)$ with $H(z)$ and $H'(z)$ in $f(X)$ cosmology. Afterwards, based on the relationship between $X$ and $H(z)$, we can obtain $f$ as the reconstructed function of $X$.
The relation between $f(X)$ and $X$, using the reconstructed $H(z)$ results for P-BAO only and DESI + P-BAO from Fig.~\ref{fig:H(z) data}, are presented in Fig.~\ref{fig:fX_X}. We mention that we do not use the DESI only result to obtain the reconstruction, since the $H(z)$ at low redshift does not behave very efficient. And we find the reconstruction results indicate $f(X)$ beyond the standard $\Lambda$CDM. We know that as the universe evolves, the absolute value of $R$, $T$ or $Q$ gradually decreases, which implies that in the late-time universe we can always perform a polynomial expansion of the gravitational   actions $f(X)$, re-expressing them as a sum of different series. However, such a description in the late-time universe is only an effective description of the original action  \cite{Oikonomou:2020oex}. In order to fit the reconstructed results of $f(X)$, we use the function form
\begin{equation}
\label{FXpaeametrization}
 F(X)/X_0=A+BX/X_0+CX^2/X_0^2 ~,
\end{equation}
where $F(X)=f(X)-X$ characterizes the derivation from general relativity, and $A,B,C$ are dimensionless parameters with $X_0$ represents the value of $X$ at current time. Finally, in Table~\ref{table:best fit value f} we provide the parameter values for different metric-affine theories and different datasets. As we can see, in all cases, the quadratic deviation from $\Lambda$CDM scenario is mildly favoured by the data.

\begin{table}[htbp]
\centering
\caption{The best-fit parameter values for the modified gravity parametrization \eqref{FXpaeametrization}, namely for $f(R)$, $f(T)$, $f(Q)$ gravity with quadratic corrections. }
\begin{tabular}{ccccc}
\hline
    Model
&\multicolumn{2}{c}{$f(R)$}
&\multicolumn{2}{c}{$f(T)$ or $f(Q)$}
\\
    \cline{1-5}
    Data&P-BAO& DESI + P-BAO&P-BAO& DESI + P-BAO \\
\hline
    $A$ &$-0.601$&$-0.531$ &0.808 &0.791  \\
    $B$ &0.0342&0.00782 &$-0.0848$ &$-0.0833$  \\
    $C$ &0.00391&0.00554 &0.00261 &0.000916 \\
\hline
\end{tabular}
\label{table:best fit value f}
\end{table}

%\section{Conclusion}
%\label{sec:conclusion}

\section{Conclusion}
The latest cosmological data released by DESI collaboration provides new insights for the exploration of the universe. In this work, we use the Hubble parameter data provided by DESI BAO and previous BAO observations to reconstruct the cosmological evolution of dynamical dark energy using Gaussian process, which indicates a quintom-B dynamics for dark energy. Then we realize this scenario within modified gravity theories and reconstruct the corresponding action functions under the $f(R)$, $f(T)$, and $f(Q)$  frameworks.

As a first step we reconstruct the Hubble parameter $H(z)$ and the EoS parameter $w(z)$ for dynamical dark energy. We find that due to the lack of low-redshift information, the five BAO data points from DESI alone are insufficient to provide a complete picture of cosmic evolution. Additionally, the value of DESI data at $z=0.51$ is beyond the $1\sigma$ allowed regions of the reconstructed $H(z)$ function. In particular, it is $2.44\sigma$ and $2.42\sigma$ away from the P-BAO only and DESI + P-BAO result, respectively. Interestingly, both P-BAO only and DESI + P-BAO datasets indicate that $w$ exhibits a quintom-B behavior, crossing $-1$ from phantom to quintessence regime. The inclusion of data from DESI shifts the crossing point of $w$ towards a higher redshift, namely from $1.80$ to $2.18$. The best fit function of the reconstructed $w$ is also given. In order to explain such a quintom-B behavior, we choose the metric-affine modified gravity theory. Particularly, we derive the iterative relationship of the function $f$ with respect to $z$. Subsequently, the corresponding functions $f(R)$, $f(T)$, and $f(Q)$ can be obtained using the reconstruction results of $H(z)$ and its high-order derivatives. Furthermore, we provide the best fit functions, and in all cases the quadratic deviation from $\Lambda$CDM diagram is mildly favored. We conclude that these modified gravity theories can yield the dynamical dark energy scenarios inclined by BAO.

It has been 20 years since the conception of quintom dark energy was first proposed \cite{Feng:2004ad}. This nontrivial phenomenon indicate the potentially dynamical nature of the late-time cosmic acceleration which renew the understanding about our universe. Now the recent DESI data release seems to hint on the quintom-B behavior and challenge the $\Lambda$CDM paradigm. While accumulated observational data is expected to bolster the corresponding confidence level, this magnificent phenomenon already pave the way for observational tests of the quintom-B theoretical framework. Modified gravity or other possible theories as alternative mechanisms hold promise for being tested as well. Although current research is still far from conclusively deciding the nature of gravitational theory, our work fosters a bridge for future precise cosmological observations and theoretical mechanisms.
\\

\section*{Conflict of interest}
The authors declare that they have no conflict of interest.

%\section*{Acknowledgments}.
\begin{acknowledgments}
We are grateful to Pierre Zhang, Xinmin Zhang and Gongbo Zhao for insightful comments. This work was supported in part by the National Key R\&D Program of China (2021YFC2203100),   the National Natural Science Foundation of China (12261131497, 12003029),   CAS young interdisciplinary innovation team (JCTD-2022-20),   111 Project (B23042),   USTC Fellowship for International Cooperation, and  USTC Research Funds of the Double First-Class Initiative.
ENS acknowledges the contribution of the LISA CosWG and the COST Actions CA18108 ``Quantum Gravity Phenomenology in the multi-messenger approach" and CA21136 ``Addressing observational tensions in cosmology with systematics and fundamental physics (CosmoVerse)".
Kavli IPMU is supported by World Premier International Research Center Initiative (WPI), MEXT, Japan.
\end{acknowledgments}

\section*{Author contributions}
Yi-Fu Cai conceived the idea. He also initiated this study with all other authors. Yuhang Yang and Xin Ren conducted numerical calculations and analyzed physical results. Qingqing Wang, Zhiyu Lu, and Dongdong Zhang helped analyze the DESI data. Yuhang Yang, Xin Ren, Qingqing Wang, Yi-Fu Cai, and Emmanuel N. Saridakis wrote the manuscript. Emmanuel N. Saridakis provided many valuable suggestions on this work. All authors discussed the results together.

\begin{widetext} 
\appendix
\section{DESI and BAO data}
\label{sec:appa}

\begin{table}[htbp]
\begin{center}
\caption{A list of BAO datasets used in this work, namely the values of $H(z)$ (in units of $\rm km$ $\rm s^{-1}Mpc^{-1}$) and their errors $\sigma_{\rm{H}}$ at redshift z.}
\begin{tabular}{c|c|c|c|c}
\hline
Survey &Index   & $z_{\rm eff}$  & $H(z)+\sigma_{\rm H}$             & Reference\\
\hline
{}    &1    & $0.51$         & $97.21\pm 2.83$    &\\
{}    &2    & $0.71$         & $101.57\pm 3.04$   & \\
DESI  &3    & $0.93$         & $114.07\pm 2.24$   & \cite{DESI:2024mwx} \\
{}    &4    & $1.32$         & $147.58\pm 4.49$   & \\
{}    &5    & $2.33$         & $239.38\pm 4.80$   & \\
\hline
{}    &6    & $0.24$         & $79.69\pm 2.99$    & \cite{Gaztanaga:2008xz} \\
{}    &7    & $0.30$         & $81.70\pm 6.22$    & \cite{Oka:2013cba}    \\
{}    &8    & $0.31$         & $78.17\pm 6.74$    & \cite{BOSS:2016zkm}\\
{}    &9    & $0.34$         & $83.17\pm 6.74$    & \cite{Gaztanaga:2008xz}\\
{}    &10    & $0.35$         & $82.70\pm 8.40$    & \cite{Chuang:2012qt} \\
{}    &11    & $0.36$         & $79.93\pm 3.39$    & \cite{BOSS:2016zkm}\\
{}    &12    & $0.38$         & $81.50\pm 1.90$    & \cite{BOSS:2016wmc}\\
{}    &13    & $0.40$         & $82.04\pm 2.03$    & \cite{BOSS:2016zkm}\\
{}    &14    & $0.43$         & $86.45\pm 3.68$    & \cite{Gaztanaga:2008xz} \\
{}    &15    & $0.44$         & $82.60\pm 7.80$    & \cite{Blake:2012pj}   \\
Previous &16 &  $0.44$         & $84.81\pm 1.83$    &\cite{BOSS:2016zkm}\\
BAO   &17    & $0.48$         & $87.79\pm 2.03$    &\cite{BOSS:2016zkm} \\
{}    &18    & $0.56$         & $93.33\pm 2.32$    & \cite{BOSS:2016zkm}\\
{}    &19    & $0.57$         & $87.60\pm 7.80$    &\cite{Chuang:2013hya} \\
{}    &20    & $0.57$         & $96.80\pm 3.40$    & \cite{BOSS:2013rlg}   \\
{}    &21    & $0.59$         & $98.48\pm 3.19$    & \cite{BOSS:2016zkm}\\
{}    &22    & $0.60$         & $87.90\pm 6.10$    & \cite{Blake:2012pj}  \\
{}    &23    & $0.61$         & $97.30\pm 2.10$    &\cite{BOSS:2016wmc}\\
{}    &24    & $0.64$         & $98.82\pm 2.99$    &\cite{BOSS:2016zkm}\\
{}    &25    & $0.978$        & $113.72\pm 14.63$  &\cite{eBOSS:2018yfg}\\
{}    &26    & $1.23$         & $131.44\pm 12.42$  &\cite{eBOSS:2018yfg}\\
{}    &27    & $1.48$         & $153.81\pm 6.39$    &\cite{eBOSS:2020uxp}\\
{}    &28    & $1.526$        & $148.11\pm 12.71$  &\cite{eBOSS:2018yfg}\\
{}    &29    & $1.944$        & $172.63\pm 14.79$  & \cite{eBOSS:2018yfg}\\
{}    &30    & $2.30$         & $224\pm 8$  &\cite{BOSS:2012gof}\\
{}    &31    & $2.36$         & $226.0\pm 8.00$    & \cite{BOSS:2013igd} \\
{}    &32    & $2.40$         & $227.8\pm 5.61$  &\cite{BOSS:2017uab}\\
\hline
\end{tabular}
\label{table:BAO data}
\end{center}
\end{table}

In this Appendix, and in particular in Table~\ref{table:BAO data}, we provide the Hubble parameter $H(z)$ used in this article, obtained from DESI and previous BAO observations like SDSS and WiggleZ.

\section{Reconstruction method}
\label{sec:appb}

In this Appendix we provide the details of the reconstruction procedure. In the case of $f(R)$ gravity we use   the   approximation \begin{align}
    f_{R} &\equiv \frac{d f(R)}{d R}=\frac{d f / d z}{d R / d 
z}=\frac{f^{\prime}}{R^{\prime}} \nonumber
\\
f^{\prime}(z) &\approx \frac{f(z+\Delta z)-f(z-\Delta z)}{2\Delta z}\\
f^{\prime \prime}(z) &\approx \frac{f(z+\Delta z)-f(z)+f(z-\Delta z)}{\Delta z^2} ~, \nonumber
\end{align}
where
$f_{R}$ can be represented by $f(z)$ and $H(z)$. Furthermore, we can 
extract the recursive relation between the $f(z_{i+1})$, $f\left(z_{i}\right)$ and $f(z_{i-1})$ as
\begin{equation}
    \begin{aligned}
        f(z_i+\Delta z)=\frac{(\frac{1}{2}+\frac{6H^2(-1-z)}{R'\Delta z^2})f(z_i)+\left[ \frac{3H^2-\rho_m}{2R'\Delta z} + \frac{R}{4R'\Delta z} - \frac{3H^2(-1-z)}{R'\Delta z^2} +\frac{3H^2R''(-1-z)}{2R^{'2}\Delta z} \right]f(z_i-\Delta z)}{\frac{3H^2-\rho_m}{2R'\Delta z} + \frac{R}{4R'\Delta z} + \frac{3H^2(-1-z)}{R'\Delta z^2} -\frac{3H^2R''(-1-z)}{2R^{'2}\Delta z}}
    \end{aligned}
    \label{eq:fR rec}
\end{equation}
where $H$ is the value of the reconstructed $H(z)$ at $z_i$, $\Delta z=z_{i+1}-z_i=z_i-z_{i-1}$, and $\rho_m=\Omega_{m0}\times 3H_0^2(1+z_i)^3$.

In the case of $f(T)$ gravity   we perform the similar approximation  
\begin{align}
 F_{T} &\equiv \frac{d F(T)}{d T}=\frac{d F / d z}{d T / d z}=\frac{F^{\prime}}{T^{\prime}}
\\
F^{\prime}(z) &\approx \frac{F(z+\Delta z)-F(z)}{\Delta z} ~,
\label{eq:approx1}
\end{align}
and thus we can acquire \cite{Cai:2019bdh, Ren:2021tfi}
\begin{equation}
 F\left(z_{i}+\Delta z \right)  =F\left(z_{i}\right) +6\Delta z \frac{H^{\prime} \left(z_{i}\right)}{H\left(z_{i}\right)}   \cdot \left[H^{2} \left(z_{i}\right) -H_{0}^{2} \Omega_{m 0} \left(1+z_{i} \right)^{3} +\frac{F\left(z_{i}\right)}{6}\right] ~.
\label{eq:fT rec}
\end{equation}
Finally, as mentioned above, for $f(Q)$ gravity we can change $T$ to $Q$, since they share the same background evolution under the coincident gauge in FRW geometry.

\end{widetext}  

\bibliographystyle{elsarticle-num}
\bibliography{DESI}

\end{document}